\documentclass[12pt]{article}
\usepackage{epsfig}
\begin{document}

\title{Nucleon QCD sum rules with the radiative corrections}

\author{V. A. Sadovnikova, E. G. Drukarev, and M. G. Ryskin\\
Petersburg Nuclear Physics Institute, \\
Gatchina, St.~Petersburg 188300,
Russia}
\date{}
\maketitle

\begin{abstract}

QCD sum rules for the nucleon are considered in complex $q^2$
plane with inclusion of the radiative corrections of the order
$\alpha_s$. It is shown that the radiative corrections affect
mainly the residue $\lambda^2$ of the nucleon pole. Their influence
on the value of the nucleon mass is much smaller. Following the
ideas of Ioffe and Zyablyuk we expand the analysis to complex
values of $q^2$. This provides a more stable solution.
Varying the weights of the contributions of different dimensions
by changing the value of the angle in the complex plane we find the
value of the six-quark condensate which insures the best consistency
of the right hand sides and left hand sides of the sum rules.
The corresponding value of the six-quark condensate appears to be
only about 10\% smaller then the one, provided by the factorization
approximation.
The value of the four-quark condensate also appears to be close to the
one, corresponding to the factorization assumption. The role of the
gluon condensate and its possible values are discussed.
\end{abstract}

\begin{center}
I. INTRODUCTION
\end{center}

The QCD sum rules (SR) method invented by Shifman  et al. \cite{1}
succeeded in expressing the static properties of the hadrons in vacuum
in terms of the expectation values of QCD operators. The SR approach
was employed by Ioffe  et al. \cite{2,3} in the case of nucleons.
The basic point is the dispersion relation for the polarization
operator
\begin{equation}
\Pi(q)\ =\ i\int d^4x e^{i(qx)}\,\langle 0|Tj(x)\bar j(0)|0\rangle
\end{equation}
of the local operator $j(x)$ carrying the quantum numbers of the
nucleon. The dispersion relation is considered at large values of
$|q^2|$, while $q^2<0$, where it can be presented as a power series of
$q^{-2}$ (with the QCD condensates being the coefficients of the
expansion). This presentation is known as the operator product expansion
(OPE) \cite{4}. At these values of $q^2$ the polarization operator can
be presented also as a power series of the QCD coupling constant
$\alpha_s$.

Several lowest order OPE terms have been obtained in \cite{2,3}. The
leading contribution of the order $q^4\ln q^2$ comes from the free
three-quark loop. The higher terms contain the expectation values
$$
\langle 0|\bar qq|0\rangle  ,\ \langle 0|\frac{\alpha_s}\pi\,
G^a_{\mu\nu}G^a_{\mu\nu}|0\rangle ,\ \langle 0|\bar qq\bar qq|0\rangle ,
$$
 etc., with $q$ and $G^a_{\mu\nu}$ standing for the quark
operators and for the gluon field tensor. The analysis of \cite{3}
included also the most important radiative corrections, in which the
coupling constant $\alpha_s$ is enhanced by the ``large logarithm"
$\ln q^2$. The corrections $(\alpha_s\ln q^2)^n$ have been included to
all orders for the leading OPE terms \cite{3,5}. This approach
provided good results for the nucleon mass \cite{3} and for the other
characteristics of nucleons \cite{6}.

There are several traditional points of SR analysis.
Following
\cite{3} one writes dispersion relations
\begin{equation}
\Pi^i(q^2)\ =\ \frac1\pi \int \frac{\mbox{Im }\Pi^i(k^2)}{k^2-q^2}\,
dk^2
\end{equation}
$(i=q,I)$ for the components of the polarization operator
\begin{equation}
\Pi(q)\ =\ q_\mu\gamma^\mu\Pi^q(q^2)+I\Pi^I(q^2)
\end{equation}
with $\gamma_\mu$ and $I$ standing for the Dirac and unit matrices.
The result of OPE on the left-hand side (LHS) of Eq.~(2),
is equalled to the hadronic contribution on the
right hand side (RHS), which is approximated by the
standard ``pole+continuum" model. The
 spectral density Im$\,\Pi^i(k^2)$ is approximated by the standard
``pole+continuum" model
\begin{eqnarray}
&& \hspace*{-0.5cm}
\frac1\pi\mbox{ Im }\Pi^i(k^2)\ =\ \lambda^2_N\delta(k^2-m^2)
\nonumber\\
&&+\ \frac1\pi\,\theta(k^2-W^2)\mbox{Im }\Pi^{i\,\rm OPE}(k^2).
\end{eqnarray}
Thus, the position of the lowest laying pole $m$, the residue
$\lambda^2_N$, and the continuum threshold $W^2$ are the unknowns of the
SR equations. The standard Borel transform is usually carried out.

However, inclusion of the lowest order radiative corrections beyond the
logarithmic approximation made the situation somewhat more complicated.
A numerically large coefficient of the lowest radiative correction to
the leading OPE of the polarization operator (1) was obtained in
\cite{7}. A more consistent calculation \cite{8} provided this
coefficient to be about 6. Thus, the radiative correction increases
this term by about 50\% at $|q^2|\sim1\rm\,GeV^2$, which are actual for
the SR analysis. This ``uncomfortably large" correction is often
claimed as the most weak point of the SR approach \cite{9}.

The radiative corrections of the order $\alpha_s\ln q^2$ and $\alpha_s$
for the contributions up to $q^{-2}$ have been calculated by
Ovchinnikov  et~al. \cite{10}. These results were used for the
calculation of the parameters $m,\lambda^2$ and $W^2$ by finite energy
sum rules method (FESR) \cite{KP}. Later the influence of the radiation
corrections on the values of the nucleon mass was studied by Borel SR
technique at fixed values of $W^2$ and $\lambda^2$, taken from FESR
calculations \cite{HS}. Recently the QCD SR with the radiative
corrections included have been used for determination of the
value of quark scalar condensate \cite{BL}. However, the analysis of
the role of the radiative corrections performed totally
in framework of the Borel transformed SR have
not been done until now.  This is done in the present paper.  We
investigate also the sensitivity of the nucleon characteristics to the
variation of the values of the condensates of higher dimensions.

These two points are connected with each other. While the condensate
$\langle 0|\bar qq|0\rangle $ of dimension $d=3$ is known with good
accuracy, there is no way to obtain the four- and six-quark condensates
$(d=6,9)$ from an experiment. The calculation of these expectation
values requires some additional assumptions.
As it stands now, the only
approach used in the calculations is the factorization hypothesis, which
assumed the domination of the vacuum states in the sum over the
intermediate states \cite{1}. In this approximation expectation
values of the products of four and six quark operators are
expressed in terms of the scalar operator $\langle 0|\bar
qq|0\rangle$.
However, the correction of the order
$\alpha_s\ln(q^2/\mu^2)$ to this condensate differs from the sum of
such corrections to the condensates $\langle 0|\bar qq|0\rangle$
\cite{10}. Thus, the factorization hypothesis should be combined with
the definition of the normalization point (scale) at which it is
assumed to be true.

We study also the dependence of the nucleon parameters on the value of
the gluon condensate $\langle 0|\frac{\alpha_s}\pi G_{\mu\nu}G_{\mu\nu}
|0\rangle $ $(d=4)$. There are indications that this matrix element can
be smaller \cite{11} or larger \cite{12} than the standard value
obtained in \cite{13}.

Analysis of the SR at complex values of $q^2=|q^2|e^{i\varphi}$ was
employed first by Ioffe and Zyablyuk \cite{14} for investigation of
hadronic $\tau$-decay. Varying the value of angle $\varphi$ in complex
plane of $q^2$ one can change the weights of different OPE
contributions. Some of the terms can be eliminated in such a way.
By using this technique one can expect to obtain more reliable and
stable results. The hadron parameters should not depend on $\varphi$.
We apply this approach for the nucleon channel.

We recall the original form of QCD SR in Sect.~II. Corrections of the
order $\alpha_s$ are included in Sect.~III. The corresponding equations
in complex $q^2$ plane and the analysis are presented in Sect.~IV.

\begin{center}
II. QCD SUM RULES AT REAL VALUES OF $q^2$
\end{center}

For the specific calculations we use the ``current" $j$, which enters
Eq.~(1) in the form \cite{2}
\begin{equation}
j(x)\ =\ u^a(x)C\gamma_\mu u^b(x)\gamma_5\gamma^\mu d^c(x)
\varepsilon_{abc}\ ,
\end{equation}
with $u$ and $d$ standing for the quark fields. $C$ is the charge
conjugation operator, while $a,b,c$ are the color indices. The lowest
OPE terms of the operators $\Pi^q(q^2)$ and $\Pi^I(q^2)$ introduced by
Eq.~(3) can be presented as
\begin{equation}
\Pi^q = A_0+A_4+A_6+A_8, \quad \Pi^I=B_3+B_7+B_9\ .
\end{equation}
Here the lower indices show the dimensions of the condensates,
contained in the corresponding terms, $A_0$ is the contribution of the
free quark loop.

The leading OPE terms $A_0$, $A_4$ and $B_3$ contain
divergent integrals. It is sufficient \cite{2} to carry out
regularization in the simplest way,  i.e., just by introducing an
ultraviolet cutoff $C_u$. Direct calculation with all the radiative
corrections being neglected provides \cite{2}
\begin{eqnarray}
&& A_0=-\frac{Q^4}{64\pi^4}\ln\left(\frac{Q^2}{C^2_u}\right),
\nonumber\\
&& A_4=-\frac1{32\pi^2}\ln\left(\frac{Q^2}{C^2_u}\right)
\langle0|\frac{\alpha_s}\pi\,G^2|0\rangle\ ,
\nonumber\\
&& B_3=-\frac{Q^2}{4\pi^2}\ln\left(\frac{Q^2}{C^2_u}\right)
\langle0|\bar qq|0\rangle\
\end{eqnarray}
with the standard notation $G^2=G^a_{\mu\nu}G^a_{\mu\nu}$, while
$Q^2=-q^2>0$.

The condensates of the lowest dimensions $d=3,4$, i.e.,
$\langle0|\bar qq|0\rangle$ and $\langle0|\frac{\alpha_s}\pi
G^2|0\rangle$ can be obtained from the Gell-Mann--Oakes--Renner
relation \cite{15} and from the meson sum rules \cite{13}.
The higher OPE terms $A_6$ and $B_9$ contain the expectation values of
the products of four and six quark operators. There are neither
experimental no rigorous theoretical data on the four-quark
condensates of the general form
$\langle 0|\bar q\Gamma_1q\bar q\Gamma_2q|0\rangle$,
with $\Gamma_{1,2}$ acting on Lorentz and color indices. The same
refers to the six-quark condensates. The standard approach is the
factorization approximatiom \cite{1} expressed by equation
\begin{eqnarray*}
&&\hspace*{-0.5cm}
\langle 0|\bar q\Gamma_1q\bar q\Gamma_2q|0\rangle\ =
\\
&& =\ N^{-2}[({\rm Tr} \Gamma_1{\rm Tr}\Gamma_2) -
{\rm Tr}(\Gamma_1\Gamma_2)]
\langle 0|\bar qq|0\rangle^2
\end{eqnarray*}
with $q$ standing for $u$ and $d$ quarks and $N=12$. This equation
presents all the four-quark condensates in terms of the scalar
expectation values $\langle 0|\bar qq|0\rangle$. However all Lorentz
structure with $\Gamma_1=\Gamma_2$ contribute due to the second term in
brackets on the RHS of the latter equation
(several examples are presented in \cite{1}).  Similar relation
can be written for the six-quark condensate. Also, the
factorization approximation for the quark-gluon condensate,
which enteres the expression for $B_7$ provides
$$
\langle0|\bar q\,\frac{\alpha_s}\pi\,G^2q|0\rangle\ =\
\langle0|\bar
qq|0\rangle \langle0|\,\frac{\alpha_s}\pi\,G^2|0\rangle\ .
$$
Thus, the higher order terms are
\begin{eqnarray}
&& A_6=\frac2{3Q^2}\,(\langle0|\bar qq|0\rangle)^2\ ,
\nonumber\\
&&A_8=-\frac1{6Q^4}\,
\mu^2_0(\langle0|\bar qq|0\rangle)^2,
\nonumber\\
&&
B_7=\frac1{12Q^2}\, \langle0|\bar qq|0\rangle
\langle0|\frac{\alpha_s}\pi G^2|0\rangle\ ,
 \nonumber\\
&& B_9\ =\ 32\pi^2\frac{17}{81}\,\frac{\alpha_s}{\pi}\,\frac1{Q^4}\,
(\langle0|\bar qq|0\rangle)^3\ .
\end{eqnarray}

Here the terms $A_6$,  $B_7$ and $B_9$ are presented in
the factorization approximation. In $A_8$ we have taken into account
that the sum rules for the baryon resonances provide \cite{BI}
\begin{equation}
\langle0|\bar qq\bar
q\,\frac{\alpha_s}\pi\,G^a_{\mu\nu}\,\frac{\lambda^a}2\,
\sigma_{\mu\nu}q|0\rangle\ =\ \mu^2_0(\langle0|\bar qq|0\rangle)^2
\end{equation}
with $\mu^2_0\approx0.8\,\rm GeV^2$.

In the analysis, carried out in \cite{3}, the most important radiative
corrections of the order $\alpha_s\ln q^2$ have been included. These
contributions were summed to all orders of $(\alpha_s\ln q^2)^n$, being
expressed in terms of the factor
\begin{equation}
L\ =\ \left(\frac{\ln Q^2/\Lambda^2}{\ln\mu^2/\Lambda^2}\right)^{4/9},
\end{equation}
with $\Lambda\approx150$ MeV is the QCD scale, while $\mu$ is the
normalization point. It was assumed in \cite{3} that the physical
characteristics are normalized at $\mu=500$~MeV. The power 4/9 reflects
the so-called anomalous dimension.

The next traditional point is the Borel transform
\begin{eqnarray}
&& \hat B\ =\ \frac{(Q^2)^{n+1}}{n!}\left(-\frac d{dQ^2}\right)^n,
\\
&& Q^2=-q^2\to\infty, \quad n\to\infty, \quad \frac{Q^2}n=M^2 ,
\nonumber
\end{eqnarray}
converting a function of $q^2$ into the function of $M^2$. It removes
the terms depending on $C_u$ in the expressions provided by Eq.~(7),
and also makes the ``pole+continuum" model more reliable,
suppressing the contributions of the heavier states by the
factor $e^{-k^2/M^2}$. Actually, in
order to deal with the values of the order of unity (in GeV units), it
is convenient to use the operator $B^*=32\pi^4\hat B$.

After these manipulations the Borel transformed SR are \cite{3}
\begin{equation}
{\cal L}^q(M^2)=R^q(M^2), \quad {\cal L^I}(M^2)=R^I(M^2)
\end{equation}
with
\begin{equation}
{\cal L}^q=\tilde A_0+\tilde A_4+\tilde A_6+\tilde A_8, \quad
{\cal L}^I=\tilde B_3+\tilde B_7+\tilde B_9\ .
\end{equation}
Here, as well as in Eq. (6), the lower indices show the dimensions of
the condensates. The terms on the RHS of Eqs. (13)
are
\begin{eqnarray}
&& \tilde A_0=\frac{M^6E_2}L ,
\, \tilde A_4=\frac{bM^2E_0}{4L},
\, \tilde A_6=\frac43\,a^2L,
\nonumber\\
&&
\tilde A_8=-\frac13\,\frac{\mu^2_0}{M^2}\,a^2, \quad
\tilde B_3=2aM^4E_1,
\nonumber\\
&&\tilde B_7=-\frac{ab}{12},
\quad
\tilde B_9=\frac{272}{81}\frac{\alpha_s}\pi\frac{a^3}{M^2}
\end{eqnarray}
with the notations $E_i=E_i(W^2/M^2), i=0,1,2$ and
\begin{equation}
a\!=\!-(2\pi)^2\langle0|\bar qq|0\rangle \mbox{ and }
b\!=\!(2\pi)^2\langle0|\frac{\alpha_s}\pi G^2|0\rangle .
\end{equation}
The conventional values of these condensates at $\mu=0.5\,$GeV are
$a=0.55\rm\,GeV^3$ and $b=0.50\rm\,GeV^4$. The functions
\begin{eqnarray}
&& \hspace*{-1cm} E_0(x)=1-e^{-x},\quad  E_1(x)=1-(1+x)e^{-x},
\nonumber\\
&& E_2(x)\ =\ 1-\left(1+x+\frac{x^2}2\right)e^{-x}
\end{eqnarray}
account for the contribution of the continuum, $E_i(x)\to1$ for
$W^2\to\infty$. Thus, the RHS of Eqs.~(12)
\begin{equation}
R^q(\!M^2)=\lambda^2e^{-m^2\!/M^2}, \
R^I(\!M^2)=m\lambda^2e^{-m^2\!/M^2}
\end{equation}
describe only the contributions of the nucleon pole with
$\lambda^2=32\pi^4\lambda_N^2$.

The LHS of Eq. (12), being calculated as OPE series,
becomes the better approximation, while the value of $M^2$ increases.
On the other hand, ``pole+continuum" model for the RHS becomes
increasingly accurate for smaller values of $M^2$. The interval
\begin{equation}
0.8\mbox{ GeV}^2<M^2<1.4\mbox{ GeV}^2,
\end{equation}
where both approximations are expected to be true was found in
\cite{3}. The values of the parameters $m,\lambda^2$ and $W^2$ have
been obtained by minimization of the functional
\begin{equation}
f_1(M^2)\ =\ \sum_{i=q,I}\left(\frac{{\cal L}^i(M^2)-R^i(M^2)}{{\cal
L}^i (M^2)}\right)^2
\end{equation}
in the interval (18), providing
\begin{eqnarray}
&& m=0.931\mbox{ GeV }, \quad \lambda^2=1.86\mbox{ GeV}^6,
\nonumber\\
&& W^2\ =\ 2.09\mbox{ GeV}^2.
\end{eqnarray}
Account of the anomalous dimensions is not very important. Putting
$L=1$ one can find
\begin{eqnarray}
&& m=0.930\mbox{ GeV},\ \lambda^2=1.79\mbox{ GeV}^6,
\nonumber\\
&& W^2\ =\ 2.00\mbox{ GeV}^2.
\end{eqnarray}

Note that straightforward using of this approach can provide
misleading results. The solution of (19) corresponding to  minimization
appears to be unstable with respect to the values of the QCD
condensates.  Moreover, sometimes the functional (19) may have
few different minima. This is illustrated by Fig.~1,
where the solution corresponding to the second minimum is
shown by the dashed line. For the values of $b$ which do not
exceed strongly the conventional value $b=0.50\rm\,GeV^4$ the best
$\chi^2$ fitting is provided by the first solution shown by
the solid line. However, for larger values of
$b$ the minimization procedure makes us to jump to the second
solution with smaller value of the nucleon mass $m=0.6\rm\,GeV$.
Consider,
for example, the SR with somewhat larger value of the gluon condensate.
At $b=0.65$ (which is 30\% larger than the conventional value)
minimization of $f_1$ (19) provides much smaller values
$m\approx0.6\,$GeV and $W^2=1\rm\,GeV^2$--see Fig.~1.
Such solution cannot be treated as a physical one because of
the small value of $W^2$.  Indeed, the contribution of the continuum
(treated approximately) exceeds the contribution of the pole, which is
treated exactly. Thus the ``pole+continuum" model for the RHS of the
sum rules has no sense.  Although the accuracy of the physical solution
does not change much with $b$, it will not be noticed by $\chi^2$
minimization procedure since unphysical solution is even more accurate
--- see Fig.~1. The unphysical solution with small values of $m$ and
$W^2$ can be traced by successive inclusion of the condensates of
higher dimension. Including only the condensates of dimensions $d=3,4$
we find a trivial solution $m=0$, $\lambda^2=0$, $W^2=0$.  Inclusion of
the condensate with $d=6$ still keeps $m=0$,$W^2=0$, but
$\lambda^2=\frac43a^2=0.4\rm\,GeV^6$. Inclusion of the higher
condensates provides small nonzero values of $m$ and $W^2$.

To avoid this situation we add analysis of the functions
\begin{eqnarray}
&&
m^2_1(M^2)=\frac{M^4}{{\cal L}^q(M^2)}\,\frac{d{\cal L}^q(M^2)}{dM^2}\,,
\nonumber\\
&& m^2_2(M^2)=\frac{M^4}{{\cal L}^I(M^2)}\,\frac{d{\cal L}^I
(M^2)}{dM^2}\,,
\nonumber \\
&& m_3(M^2)\ =\ \frac{{\cal L}^I(M^2)}{{\cal L}^q(M^2)}\,.
\end{eqnarray}
Note that these functions depend also on the threshold value
$W^2$. For the solution of the SR equations
(12) it should be
\begin{equation}
m^2_1(\!M^2)=m^2_2(\!M^2)=m^2_3(\!M^2)=\mbox{const}= m^2.
\end{equation}
Note different meaning of the masses $m_{1,2}$ and $m_3$. While
$m_{1,2}$ determine the position of the lowest pole,
which approximates the $q^2$ dependence of (3),
the difference
$m_3-m$ reflects rather the admixture of the negative parity state
$1/2^-$ to the nucleon.

\begin{center}
III. CORRECTIONS OF THE ORDER $\alpha_s$
AT REAL VALUES OF $q^2$
\end{center}

Inclusion of the corrections of the order $\alpha_s$ to the
polarization operator modifies the expressions on the RHS of
Eqs.~(7),(8) to
\begin{eqnarray}
A_0 &=& -\ \frac1{64\pi^4}\,Q^4\ln\frac{Q^2}{\mu^2}
\nonumber\\
&& \times\ \left(1+\frac{71}{12}
\frac{\alpha_s}\pi-\frac12\frac{\alpha_s}\pi\ln\frac{Q^2}{\mu^2}\right),
\nonumber\\
A_6 &=&\frac23\,\frac{\langle0|\bar qq\bar qq|0\rangle}{Q^2}
\left(\!\!1\!-
\frac56\frac{\alpha_s}\pi-\frac13\frac{\alpha_s}\pi\ln
\frac{Q^2}{\mu^2}\!\right)\!,
\nonumber \\
B_3 &=& - \frac{\langle0|\bar qq|0\rangle}{4\pi^2}\,Q^2\ln
\frac{Q^2}{\mu^2}\left(1+\frac32\,\frac{\alpha_s}\pi\right),
\end{eqnarray}
obtained in \cite{8,10} (see Appendix A).
The radiative corrections to the other terms are not included
since the values of the corresponding condensates are
known with poor accuracy. Strictly speaking this is true
for the four-quark condensate as well, which is obtained
in the factorization approximation only. However, we shall
assume the latter hypothesis for the four-quark condensate,
advocated recently in \cite{MS}.
Note that
the correction of the type $\alpha_s\ln(q^2/\mu^2)$ to the term $B_3$
vanishes due to the cancellation of the correction
$\frac12\frac{\alpha_s}\pi\ln\frac{Q^2}{\mu^2}$ to the condensate
$\langle0|\bar qq|0\rangle$ and that coming from the free quark system.
If the factorization approximation  would have been true for all
values of $\mu^2$, we would find
$+\frac{\alpha_s}\pi\ln\frac{Q^2}{\mu^2}$ for the last term in
brackets of the formula for $A_6$. One can see that the value of the
coefficient is another one. Thus, the factorization assumption cannot
be true for all values of $\mu^2$.
\footnote{Since
the value of $\ln(M^2/\mu^2)$ is not large and we are
studying the role of the whole $O(\alpha_s)$ corrections, and
not only the part (given by the renormgroup) which is
responsible for the scale dependence of the polarization
operator, here we keep the logarithmic $\alpha_s\ln Q^2$
contribution but, to avoid the double counting, neglect the
anomalous dimension factor $L^\gamma$.}

In Ref. \cite{10} these formulas have been used for investigation of
the nucleon parameters with the finite energy sum rules technique
\cite{KP}. It was shown that inclusion of radiative correction leads to
a noticeable reduction of the nucleon mass. Similar behavior was
obtained in the analysis carried out in \cite{HS} in framework of the
Borel transformed sum rules at fixed values of $W^2$.

Using (24), we find for the corresponding contributions to the Borel
transformed SR (see Appendix B)
\begin{eqnarray}
&& \hspace*{-0.7cm}
\tilde A_0(M^2,W^2)=M^6E_2\!\left[\!1+\frac{\alpha_s}\pi\!\left(
\!\frac{53}{12}-\ln\frac{W^2}{\mu^2}\!\right)\!\right]
\nonumber\\
&&
-\ \frac{\alpha_s}{\pi}\bigg[M^4W^2\!\left(1+\frac{3W^2}{4M^2}
\right)e^{-W^2/M^2}
   \nonumber\\
&&+\ M^6 {\cal E}\!\left(-{\rm W^2/ M^2}\right)\bigg],
\nonumber\\
&& \hspace*{-0.7cm}
\tilde A_6(M^2,W^2)\ =\ \frac43\,a^2
\\
&&\hspace*{-0.3cm}
\times \left[\!1\!-\frac{\alpha_s}\pi\!\left(\!\frac56+\frac13\!
\left(\!\ln\frac{W^2}{\mu^2}\!+{\cal E}(-{\rm
W^2/M^2})\!\right)\!\right)\!\right]\!,
\nonumber\\
&& \hspace*{-0.7cm}
\tilde B_3(M^2,W^2)\ =\ 2aM^4E_1\left(1+\frac32\,\frac{\alpha_s}\pi\right)
\nonumber
\end{eqnarray}
with  $E_i=E_i(W^2/M^2)$, $i=0,1,2$ and
$$
{\cal E}(x) =\ \sum_{n=1}\,\frac{x^n}{n\cdot n!}\ .
$$

Expression for $\tilde B_3$ coincides with that presented in
\cite{HS}. Two other ones are more general than those of \cite{HS},
being true for any value of $\mu^2$, and accounting for the
contribution of continuum in the main radiative correction
to the term of the sixth dimension. Besides that some terms in the
expression for $\tilde A_0$ differ from those in \cite {HS}.
The numerical difference is, however, not very important.

One can see that the free loop term $\tilde A_0$ suffers the largest
radiation corrections. If the threshold $W^2$ is large enough, $\tilde
A_0$ provides the leading contribution to the structure ${\cal L}^q$.
Hence, neglecting the influence of radiative corrections on the value
of $W^2$, one can see that the denominator of the function $m_3(M^2)$
(22) increases due to the radiative corrections. Thus, in agreement
with \cite{10,HS}, the latter diminish the value of nucleon mass.
However, the present analysis which includes possible modification of
all the parameters $(m,\lambda^2,W^2)$ by the radiative corrections,
provides somewhat different result.

To carry out quantitative analysis, we need to clarify the argument of
the running coupling constant
\begin{equation}
\alpha_s(k^2)\ =\ \frac{4\pi}{9\ln(k^2/\Lambda^2)}\ .
\end{equation}
In \cite{10} the radiative corrections have been obtained at
$\alpha_s=\,$const. Since the momenta in the loops corresponding to the
radiative corrections are of the order of $q$, it is reasonable to
assume $\alpha_s=\alpha_s(Q^2)$ in Eq.~(24). After the Borel transform
we obtain $\alpha_s=\alpha_s(M^2)$ in Eq.~(25). Since we are
considering $M^2$ of the order $1\rm\,GeV^2$, we can put
$\alpha_s=\alpha_s(\rm1\,GeV^2)=0.35$. Minimization of the functional
(19) provides
\begin{eqnarray}
&&\hspace*{-1cm} m=0.94\mbox{ GeV }, \quad \lambda^2=2.00\mbox{ GeV}^6,
\nonumber\\
&&  W\ =\ 1.90\mbox{ GeV}^2,
\end{eqnarray}
while for $\alpha_s=\alpha_s(M^2)$
\begin{eqnarray}
&& \hspace*{-1cm}
m=0.94\mbox{ GeV }, \quad \lambda^2=2.11\mbox{ GeV}^6,
\nonumber\\
&& W^2\ =\ 2.00\mbox{ GeV}^2.
\end{eqnarray}

In Fig. 2 we present the values of the parameters as the functions of
$\alpha_s$. One can see that the radiative corrections affect mostly
the value of the residue $\lambda^2$.
The consistency of the LHS and RHS of the SR
is shown in Fig.~3(a).

Note that in \cite{BL} the SR have been considered at fixed value
$W^2=2.5 \rm\,GeV^2$, providing somewhat larger residue  value
$\lambda^2 \approx 3$ GeV$^2$. Assuming $W^2=2.5$ GeV$^2$ in Eqs.~(12),
we would move to larger values $\lambda^2 =2.6 $GeV$^2$ in our solution.
This is close to the result of \cite{BL}, although our procedures of
inclusion of radiative corrections differ in several points.

We can look for the solution in which the discrepancy between the
values of $m_i(M^2)$ defined by Eq.~(22) is minimized. It can be
obtained by minimization of the functional
\begin{equation}
f_2(M^2)\ =\ f_1(M^2)+f_m(M^2)
\end{equation}
with
\begin{equation}
f_m(M^2)\ =\ \sum^3_i \frac{(m_i(M^2)-m)^2}{3m^2}\ ,
\end{equation}
while $f_1(M^2)$ is determined by (19).
For $\alpha_s=0$ the procedure provides $m=0.93\,$GeV,
$\lambda^2=1.82\rm\,GeV^6$, $W^2=2.02\rm\,GeV^2$, while for
$\alpha_s=0.35$ we find
\begin{eqnarray}
&& \hspace*{-1cm}
m=0.94\mbox{ GeV }, \quad \lambda^2=1.98\mbox{ GeV}^6,
\nonumber\\
&& W^2\ =\ 1.88\mbox{ GeV}^2.
\end{eqnarray}

Thus, minimization of the functionals $f_1(M^2)$ (19) and $f_2(M^2)$
(29) leads to close results.  The functions $m_i(M^2)$ are presented in
Fig.~3(b).

\begin{center}
IV. SUM RULES IN THE COMPLEX PLANE OF $Q^2$
\end{center}

Now we consider the SR at complex values of
\begin{equation}
Q^2\ =\ S^2 e^{i\varphi}
\end{equation}
with the real values of $S^2>0$ and $-\pi<\varphi<\pi$. The Borel
transform (11) will be carried out with respect to $S^2$. It is
reasonable to consider the SR for the real part of the operators
$\Pi^i(q^2)$ (3).

\begin{center}
A. Possible values of $\varphi$
\end{center}

Note first that our ``pole+continuum" model makes sense only for
$\cos\varphi>0$, i.e., $-\frac\pi2<\varphi<\frac\pi2$. This becomes
clear while considering the Borel transform of contribution of the
nucleon pole to the RHS of Eq.~(2)
\begin{equation}
B^*_{S^2}\!\int\!\frac{\lambda^2_N\delta(k^2-m^2)}{m^2
+S^2e^{i\varphi}}\,
dk^2 = \lambda^2e^{-\frac{m^2}{M^2}\cos\varphi} e^{i\theta}
\end{equation}
with $\theta=\frac{m^2}{M^2}\sin\varphi$. Thus the contributions of
heavier states are suppressed with respect to the lowest one only if
$\cos\varphi>0$.

Also, the values of $\varphi$ should not be too close to $\pm\pi/2$.
This is because the leading OPE term $B_3$ obtains the factor
$\cos\varphi$ in the complex plane, while the higher term $B_9$ is
multiplied by $\cos2\varphi$. Hence, the convergence of OPE series
becomes worse at $\varphi$ close to $\pm\pi/2$. That's why we focus on
the values
\begin{equation}
0\ \le\ \varphi\ \le\ \frac\pi4\ .
\end{equation}
Analysis at negative values of $\varphi$ will not provide new data,
since all the functions involved are the even functions of $\varphi$.

\begin{center}
B. Basic equations in the complex plane
\end{center}

Following the previous discussion, we must present the Borel
transformed dispersion relations for the real parts of the operators
$\Pi^q(q^2)$ and $\Pi^I(q^2)$. For the Borel transforms of the
contributions (7), (8) we find
\begin{eqnarray}
&& \hspace*{-0.5cm}
\hat BA_0=\frac{M^6e^{2i\varphi}}{32\pi^4}, \quad \hat
BA_4=\frac{M^2}{32\pi^2}\langle0|\frac{\alpha_s}\pi\,G^2|0\rangle\,,
\nonumber\\
&&
\hat BA_6=\frac23 e^{-i\varphi}\langle0|\bar qq\bar qq|0\rangle\,,
\nonumber \\
&& \hspace*{-0.5cm}
\hat BA_8=\frac{-1}{6M^2}\, e^{-2i\varphi}\langle0|\bar qq \bar q
\frac{\alpha_s}\pi\,G^a_{\mu\nu}\,\frac{\lambda^a}2\,\sigma_{\mu\nu}
q|0\rangle\ ,
\nonumber \\
&& \hat BB_3=-\frac{M^4}{4\pi^2}\,e^{i\varphi}\langle0|\bar qq|0
\rangle\ ,
\nonumber\\
&& \hat BB_7=\frac{e^{-i\varphi}}{24}\,\langle0|\bar
q\,\frac{\alpha_s}\pi G^2q|0\rangle\ ,
\nonumber \\
&& \hat BB_9=\frac{272\,e^{-2i\varphi}}{81
M^2}\,\frac{\alpha_s}{\pi^3}\, \langle0|\bar qq\bar qq\bar
qq|0\rangle\ .
\end{eqnarray}

The contributions $A_0$, $A_4$ and $B_3$ contain the quark loops, thus
requiring inclusion of the contribution of the continuum. The latter
can be evaluated as continuum, which takes the form
\begin{equation}
R_n\ =\ \int\limits^\infty_{W^2}
\frac{dk^2(k^2)^n}{k^2+S^2e^{i\varphi}}\ =\ \int\limits_{W^2_c}
\frac{dk^2_c(k^2_c)^ne^{in\varphi}}{k^2_c+S^2}
\end{equation}
with $n=0,1,2,$ and
\begin{equation}
W^2_c\ =\ W^2e^{-i\varphi}.
\end{equation}
The corresponding Borel transforms are thus
\begin{equation}
\hat BR_n\ =\ e^{in\varphi}M^2\left(1-E_n\left(
\frac{W^2_c}{M^2}\right)\right).
\end{equation}

Using these results, and including the radiative corrections (25) we
can present the SR in the form
\begin{equation}
\mbox{Re }({\cal L}^q_c e^{i\varphi})
=R^q_c, \quad \mbox{ Re }({\cal L}^I_c e^{i\varphi})=R^I_c
\end{equation}
with
\begin{eqnarray}
&& \hspace*{-1cm} R^q_c=\lambda^2
e^{-\frac{m^2}{M^2}\cos\varphi}\cos\left(\frac{m^2}{M^2}
\sin\varphi\right),
\nonumber\\
&& \hspace*{-1cm} R^I_c = m\lambda^2 e^{-\frac{m^2}{M^2}\cos\varphi}
\cos\left(\frac{m^2}{M^2}\sin\varphi\right),
\end{eqnarray}
while
\begin{eqnarray}
&&   \hspace*{-0.7cm}
{\cal L}^q_c(M^2,W^2_c)\!=\! e^{2i\varphi}\tilde A_0(M^2,W^2_c)\!
+\!\tilde A_4(M^2,W^2_c)
\nonumber\\
&&+\ e^{-i\varphi}\tilde A_6(M^2)
+e^{-2i\varphi}\tilde A_8(M^2),
\\
&& \hspace*{-0.7cm}
{\cal L}^I_c(M^2,W^2_c) = e^{i\varphi}\tilde B_3(M^2,W^2_c)
\nonumber\\
&&+\ e^{-i\varphi}\tilde B_7+e^{-2i\varphi}\tilde B_9(M^2)
\end{eqnarray}
with $W^2_c$ defined by Eq. (37), while the contributions $\tilde A_i$
and $\tilde B_i$ are given by Eq.~(25). The functions
\begin{eqnarray}
&& m^2_1(M^2)=\mbox{ Re }\frac{M^4e^{i\varphi}}{{\cal L}^q(M^2)}
\frac{d{\cal L}^q (M^2)}{dM^2}\ ,
\nonumber\\
&& m^2_2(M^2)=\mbox{ Re }\frac{M^4e^{i\varphi}}{{\cal
L}^I(M^2)} \frac{d{\cal L}^I(M^2)}{dM^2}\ ,
\nonumber\\
&& m_3(M^2)\ =\
\frac{\mbox{ Re }{\cal L}^I(M^2)}{\mbox{ Re }{\cal L}^q(M^2)}
\end{eqnarray}
generalize Eq. (22) for the complex plane, and Eq.~(23) should have
been true for the exact solution

\begin{center}
C. Dependence of hadron parameters on $\varphi$
\end{center}

Now we try to find the nucleon parameters by minimization of the
functional
\begin{equation}
f^c_1(M^2)=\sum_i\left(\frac{\mbox{Re }{\cal L}^i_c(M^2)-R^i_c(M^2)}{
\mbox{Re }{\cal L}^i_c(M^2)}\right)^2,
\end{equation}
which is generalization of Eq. (19) for the complex plane of $Q^2$.
Since the $M^2$ interval of stability may vary with
$\varphi$, we present the results for the two intervals, e.g., the
one given by Eq.~(18), and also
\begin{equation}
1.0\mbox{ GeV}^2\ <\ M^2\ <\ 1.8\mbox{ GeV}^2.
\end{equation}
The results are shown in Fig.~4. One can see that the largest
stability is obtained at $\varphi\le\frac\pi4$. Also, the nucleon
residue appears to be more sensitive to the value of $\varphi$, then
the values of the nucleon mass and of the continuum threshold $W^2$.

\begin{center}
D. Dependence on the value of six-quark condensate
\end{center}

As we said earlier, the less reliable value of the condensates involved
is that of the six-quark condensate,
which determines the contribution $\tilde B_9$ on the RHS
of Eq.~(42). While all the results obtained above are obtained in
factorization approximation, now we put
\begin{equation}
\langle0|(\bar qq)^3|0\rangle\ =  \left(\langle0|\bar
qq|0\rangle\right)^3\eta_{6q}\ ,
\end{equation}
with $\eta_{6q}=1$ under the factorization assumption and check
dependence of the parameters on $\varphi$ at various values of
$\eta_{6q}$. The results are presented in Fig.~5. One can see
that the value of the nucleon mass $m$ exhibits larger stability with
respect to changes of $\varphi$ and $\eta_{6q}$ then the values of
$\lambda^2$ and $W^2$.

At $\varphi=\frac\pi4$ the contribution of the six-quark condensate
vanishes, and the result is free from uncertainties in the value of
$\eta_{6q}$. For $\varphi=\frac\pi4$
\begin{eqnarray}
&& \hspace*{-0.5cm}
m=0.95\mbox{ GeV }, \quad \lambda^2=1.49\mbox{ GeV}^6,
\nonumber\\
&& W^2\ =\ 1.57\mbox{ GeV}^2.
\end{eqnarray}
Note that at $\varphi=\pi/4$ the contribution $\tilde A_0$ on the RHS
of Eq.~(41), corresponding to the free-quark loop also turns to zero.
This is not necessary a weak point, since the system of three free
quarks contributes rather to continuum then to the lowest laying pole.

At $\varphi=\pi/4$ the consistency of LHS and RHS of Eqs.~(39)
is not very good. It becomes much better at smaller values of $\varphi$
in the domain of stability in $(M^2,\varphi)$ plane.

\begin{center}
E. Stability of solution in $(M^2,\varphi)$ plane
\end{center}

Now we shall look for the solution which provides the best consistency
of the LHS and RHS of Eqs. (39) in certain domains of the
values of $M^2$ and $\varphi$. In Table I we present result obtained by
minimization of the functional (44). One can see that the consistency
(``$\chi^2$ per point") becomes much better if we limit the interval of
the values of the angles to
\begin{equation}
0\ \le\ \varphi\ \le\ \frac\pi8\ .
\end{equation}

The results obtained in the interval (45) are close to those found for
the interval (18). In this case ''$\chi^2$ per point" is somewhat
smaller than in the latter one. However we consider the results
obtained in traditional interval to be more reliable, since on the
upper end of (45) the contribution of continuum is not suppressed
stronger than that of the pole. Thus, the limits of stability of SR in
$(M^2,\varphi)$ plane are given by Eqs. (18) and (48) with the values
of the parameters being
\begin{eqnarray}
&& \hspace*{-0.5cm}
m=0.94\mbox{ GeV }, \quad \lambda^2=2.0\mbox{ GeV}^6,
\nonumber\\
&& W^2\ =\ 1.9\mbox{ GeV}^2.
\end{eqnarray}
Consistency of LHS and RHS of Eqs. (39) is illustrated by
Fig.~6 for the hadron parameters presented by Eq.~(49)
for several values
of $\varphi$.  In Fig.~7 we show the functions $m_i(M^2)$
defined by Eq.~(43) for several values of $\varphi$ and dependence of
the values $m_i(M^2)$ on $\varphi$ for several values
of $M^2$.  This illustrates the weak dependence of the values of the
hadron parameters on the values of $M^2$ and $\varphi$ in this region.

Note that similar results can be obtained by minimization of the
functional
\begin{equation}
f_2^c(M^2)\ =\ f_1^c(M^2)+f_m(M^2)
\end{equation}
with $f_m$ and $f_1^c$ defined by Eqs. (30) and (44). These data are
presented in Table~II .  The functions $m_i(M^2)$ are shown in Fig.~8.

\begin{center}
F. Values of the condensates of high dimension
\end{center}

Now we can try to use the SR for diminishing of the uncertainties of
the values of the QCD condensates. We put
\begin{eqnarray}
&& \hspace*{-0.5cm}
\langle0|(\bar qq)^2|0\rangle= (\langle0|\bar qq|0\rangle)^2\eta_{4q}\,,
\nonumber\\
&& \hspace*{-0.5cm}
\langle0|\frac{\alpha_s}\pi G^2|0\rangle = (2\pi)^{-2}b_0\eta_G
\end{eqnarray}
with $b_0=0.50\rm\,GeV^4$, and try to determine the values of
$\eta_{4q}$, $\eta_G$, and $\eta_{6q}$ (46) from the SR. Deviations of
the parameters $\eta_{4q}$ and $\eta_{6q}$ from unity characterize the
violation of factorization hypothesis.
 Possible deviation of parameter $\eta_G$
from unity is due to uncertainties in the value of the gluon
condensate. The terms $\tilde A_4$, $\tilde A_6$ and $\tilde B_9$ on
the RHS of Eqs. (41) and (42) obtain the factors $\eta_G$, $\eta_{4q}$
and $\eta_{6q}$ correspondingly.

Now we try to solve SR equations treating the hadron parameters
$m,\lambda^2,W^2$ and also the values $\eta_{4q}$, $\eta_{6q}$ and
$\eta_G$ as the unknowns.

\noindent
{\bf Four-quark condensate.}
Here we put $\eta_{6q}=\eta_G=1$ and try to determine the value of
$\eta_{4q}$. Minimization of (44) provides
\begin{eqnarray}
&& m=0.964\mbox{ GeV }, \quad \lambda^2=2.11\mbox{ GeV}^6,
\nonumber\\
&& W^2=1.99\mbox{ GeV}^2, \quad \eta_{4q}=0.93 .
\end{eqnarray}
Thus, the QCD sum rules prefer only small deviation from the
results of the factorization approximation.

\noindent
{\bf Six-quark condensate.}
Now we try to determine the value of $\eta_{6q}$, putting
$\eta_{4q}=\eta_G=1$. Minimization of (44) provides
\begin{eqnarray}
&& m=0.937\mbox{ GeV }, \quad \lambda^2=1.97\mbox{ GeV}^6,
\nonumber\\
&& W^2=1.90\mbox{ GeV}^2, \quad \eta_{6q}=0.90,
\end{eqnarray}
demonstrating small deviations of the value $\eta_{6q}$ from the
factorization hypothesis value $\eta_{6q}=1$.

\noindent
{\bf Gluon condensate.}
Now we fix $\eta_{4q}=\eta_{6q}=1$ and look
for the dependence of the solution on the value of
the gluon condensate. The value of the nucleon
mass appears to be rather sharp function of
$\eta_G$. For example, at $\eta_G=2$ we find
$m=0.74 \rm\,GeV$.

Note that the functions $m_3(M^2)$ (22) is much more sensitive to the
value of the gluon condensate than the ``pole masses" $m_{1,2}(M^2)$.
Variation of the parameter $\eta_G$ in the interval $0\le\eta_G\le2$
changes the values of $m_{1,2}$ by about 60\,MeV, while $m_3$ changes
by a factor 1/2.

Treating $\eta_G$ as an unknown parameter of the SR equations
we find
that minimization of  (44) takes place at
$\eta_G=2.2$
and at the
underestimated value of the nucleon mass $m\approx0.67\rm\,GeV$. On the
other hand, one needs $\eta_G \approx 1$ to obtain the value of the
nucleon mass close to the observable one.

\begin{center}
V. SUMMARY
\end{center}

We carried out the analysis of QCD sum rules for nucleons including the
lowest order radiative corrections. This is the first analysis
performed totally in framework of SR approach. We show that the
radiative corrections modify mainly the values of the nucleon residue,
while that of the nucleon mass suffers minor changes.

Our analysis was carried out for the real values of $q^2$ and in the
complex $q^2$ plane. We found the region of stability in
$(M^2,\varphi)$ plane in the latter case. Our main result is expressed
by Eq.~(49), being illustrated by Fig.~6.

We used the nucleon sum rules to clarify the values of the condensates
of high dimension.
We showed that the sum rules require the deviations
of the four-quark condensates from the factorization value to be very
small. Similar deviations of the six-quark condensates do not exceed
$10\%$. As to the uncertainties of the gluon condensate, the
values, which exceed the standard one
by $30\%$ provide the value of the nucleon mass
$m<0.9\rm\,GeV$. The
greater values of the gluon condensate may cause problems in
description of the nucleon.  A more detailed analysis of the
limitations on the gluon condensate value, coming from the nucleon sum
rules will be published elsewhere.

\begin{center}
ACKNOWLEDGMENTS
\end{center}

We thank B. L. Ioffe and A. A. Pivovarov for fruitful discussions.
The work was supported by the grant RFBR 03-02-1724-a.
We acknowledge also the partial support by the grant
RSGSS-1124.2003.02.

\begin{center}
APPENDIX A
\end{center}

Here we describe the renormalization
procedure, which leads to Eqs.~(24).
The point have not
been considered in  detail in previous publications.
Until we do not include the radiative corrections
to the quark loops (terms $A_0$, $A_4$ and $B_3$ in Eqs.~(7),~(8)),
renormalization is not important, since the contribution $Q^4
ln(C_u^2)$ is eliminated by the Borel transform.  In other words,
the Borel transform carries out the renormalization automatically.
However, the situation becomes more complicated if the corrections of
the order $\alpha_s ln Q^2$ are included.

Consider first the three-quark loop. While the radiative corrections
are neglected, we can carry out the renormalization
of the function $A_0(Q^2)$ by subtracting
the two lowest terms of the Tailor expansion
at a normalization point $Q^2=\mu^2$.
This provides for the renormalized contribution
$$
A_{0r}\ =\ -\frac1{64\pi^4}\,Q^4\ln\frac{Q^2}{\mu^2}.
\eqno{(A1)}
$$
(we omitted the terms, which will be killed by the Borel transform).
The contributions which include the radiative corrections to Eq.~(A1)
are renormalized in the same way. This leads
to the equality, presented by the upper line of Eq.~(24).

Similar procedure can be applied to the calculation of the
renormalized contributions $A_6$ and $B_3$. The result is presented
by Eq.~(24). There are no loops in the case of $A_6$. Thus, only the
radiative corrections are renormalized. There are no corrections
of the order $\alpha_s lnQ^2$ in the case of $B_3$. Hence, only
the two quark loop was renormalized. Unlike the case of $A_0$ ,
the renormalization was not important here due to the Borel transform.

\begin{center}
APPENDIX B
\end{center}

Here we calculate the contribution of the term $A_0$ determined by
Eq.~(24) to the LHS of Eq.~(12).  This means that we must calculate
the Borel transform, subtracting the contribution of the
continuum.

The renormilized contribution is
$$
A_{0r}\ =\
-\frac1{64\pi^4}\,Q^4\ln\frac{Q^2}{\mu^2}
$$
$$
\times\ \left(1+ \frac{71}{12}\,
\frac{\alpha_s}\pi-\frac12\,\frac{\alpha_s}\pi
\ln\frac{Q^2}{\mu^2}\right).
\eqno{(B1)}
$$

Employing \cite{2,HS} we find
$$
\hat B(Q^4\ln Q^2)=-2M^6,
$$
$$
\hat B(Q^4\ln^2 Q^2)=-4M^2\left(\ln M^2
-\gamma_E+\frac32\right)
\eqno{(B2)}
$$
with $\gamma_E\approx0.577$ being the Euler constant. Thus
$$
B^*A_{0r}\ =\ M^6\left(1+\frac{71}{12}\,\frac{\alpha_s}\pi\right)
$$
$$
-\ M^6
\frac{\alpha_s}\pi\left(\ln\frac{M^2}{\mu^2}-\gamma_E+\frac32\right)
\eqno{(B3)}
$$
with the last term of the RHS originating from the last term in
brackets on the RHS of Eq.~(B1). Recall that $B^*=32\pi^4\hat B$.

Following Eq. (4) we find for the contribution to the RHS of Eq.~(13)
for ${\cal L}^q$
$$
\tilde A_0(M^2,W^2)\ =\ M^6E_2\left(1+
\frac{71}{12}\frac{\alpha_s}\pi\right)
$$
$$
-\ \frac{\alpha_s}\pi\!\left(\!M^6\!\left(\!\ln\frac{M^2}{\mu^2}\!
-\gamma_E+\frac32\!\right)\!-T(M^2,W^2)\!\right),
\eqno{(B4)}
$$
with
$$
T(M^2,W^2)\ =\ \frac12\int\limits^\infty_{W^2}dk^2k^4\ln
\frac{k^2}{\mu^2}\,e^{-k^2/M^2}
\eqno{(B5)}
$$
describing the contribution of continuum, corresponding to the last
term on the RHS of Eq.~(B3).

Introducing
$$ z\ =\ \frac{W^2}{M^2}\ ,
\eqno{(B6)}
$$
and denoting $k^2=W^2x$, we evaluate
$$
T(M^2,W^2)\ =\ M^6
$$
$$
\times\
\left[\left(1+z+\frac{z^2}2\right)e^{-z}\ln
\frac{W^2}{\mu^2}+t(z)\right]
\eqno{(B7)}
$$
with
$$
t(z)\ =\ \frac{z^3}2\int\limits^\infty_1 dx\,x^2\ln xe^{-xz}.
\eqno{(B8)}
$$
Employing integration by parts three times, we obtain
$$
t(z)\ =\int\limits^\infty_1 dx\frac{e^{-xz}}x+\frac{3+z}2\,e^{-z}
$$
$$
=\ -Ei(-z)+\frac{3+z}2\,e^{-z},
\eqno{(B9)}
$$
with
$$
Ei(x)={\rm V.p.}\!\!\int\limits^x_{-\infty}\!dy\frac{e^{-y}}y\!=\!
\sum_{n=1}\!\frac{x^n}{n\cdot n!}+\gamma_E+\ln x
\eqno{(B10)}
$$
being the integral exponential function \cite{A}.
Combining Eqs. (B4), (B7)
and (B9), we come to Eq.~(25) of the text.

\newpage

{\bf Figure captions}

\noindent
FIG. 1. Dependence of solutions of Eq.~(12) on the value of the gluon
condensate. The solid lines correspond to $b$=0.50, the dashed lines
are for the case $b$=0.65.
 The nucleon parameters are related to those,
determined by Eq.~(21), i.e., $m_0=0.93$~GeV,
$\lambda^2_0=1.79$~GeV$^6$, $W^2_0=2.00$~GeV$^2$.

\noindent
FIG. 2. Dependence of ratios $m/m_0$, $\lambda^2/\lambda_0^2$,
and $W^2/W_0^2$ on the
values of $\alpha_s$.  Long-dashed line shows "$\chi^2$ per point".
Parameters $m_0$, $\lambda^2_0$ and $W^2_0$ are the same as in Fig.~1.

\noindent
FIG. 3. (a) Consistency of LHS and RHS of Eq.~(12). Hadron parameters are
given by Eq.~(28). Solid and dashed lines show the  ratios
$\frac{{\cal L}^q(M^2)}{R^q(M^2)}$ and
$\frac{{\cal L}^I(M^2)}{R^I(M^2)}$, correspondingly.
(b) The functions $m_i(M^2)$, $i$=1,2,3, defined by Eq.~(22).

\noindent
FIG. 4.
Dependence of the hadron parameters on the value of $\varphi$.
Figures (a) and (b) correspond to the values of Borel masses in the
intervals (18) and (45).
Parameters $m_0$, $\lambda^2_0$ and $W^2_0$
are the same as in Fig.~1.

\noindent
FIG. 5. Dependence of the hadron parameters $m$ (a),
$\lambda^2$ (b) and $W^2$ (c) on the value
of $\varphi$ for several values of $\eta_{6q}$. The latter
are shown by the
numbers on the lines.

\noindent
FIG. 6. Consistency of LHS and RHS of Eq.~(39) with the hadronic parameters
presented by Eq.~(49) for
$\varphi=0$ (solid line),
$\varphi=\frac{\pi}{16}$ (dashed line) and
$\varphi=\frac{\pi}8$ (dotted line).

\noindent
FIG. 7. (a)
The functions $m_i(M^2)$,
 $i$=1,2,3, for $\varphi$=0 (solid line) and
$\varphi=\frac{\pi}8$ (dashed line).
The functional $f_1(M^2)$ (19) is
minimized.  (b) Dependence of the functions $m_i(M^2)$ on the value of
$\varphi$ for $M^2$=0.8~GeV$^2$ (solid line) and for $M^2$=1.4~GeV$^2$
(dashed line).

\noindent
FIG. 8. The functions $m_i(M^2)$,
 $i$=1,2,3, for $\varphi$=0 (solid line) and
$\varphi=\frac{\pi}8$ (dashed line).
The functional $f^c_2(M^2)$ (50) is minimized.

\newpage

TABLE I. Nucleon parameters obtained by minimization\\
of the function $f_1^c$ defined by Eq.~(44).

\begin{tabular}{cccccc}
\hline
\hline
$M^2$ (GeV$^2$) & $\phi$ & $\chi^2$ & $m$ (GeV) & $\lambda^2$ (GeV$^6$) &
$W^2$ (GeV$^2$)
\\ \hline
 0.8 - 1.4 & 0-$\frac\pi4$ & 6.5 &0.93 &  1.88 &  1.84
\\
 1.0 - 1.8 & 0-$\frac\pi4$ & 2.0 &0.96 &  1.98 &  1.86
\\
 0.8 - 1.4 & 0-$\frac\pi8$ & 0.8 &0.94 &  2.02 &  1.93
\\
 1.0 - 1.8 & 0-$\frac\pi8$ & 0.4 &0.96 &  2.00 &  1.88
\\
\hline
\hline
\end{tabular}

\vspace{3cm}

 TABLE II. Nucleon parameters obtained by minimization\\
of the function $f_2^c$ defined by Eq.~(50).

\begin{tabular}{cccccc}
\hline
\hline
$M^2$ (GeV$^2$) & $\phi$ & $\chi^2$ & $m$ (GeV) & $\lambda^2$ (GeV$^6$) &
$W^2$ (GeV$^2$)
\\ \hline
 0.8 - 1.4 & 0-$\frac\pi4$ & 7.9 & 0.94 &  1.93 & 1.88 \\
 1.0 - 1.8 & 0-$\frac\pi4$ & 3.7 & 0.97 &  1.99 & 1.87 \\
 0.8 - 1.4 & 0-$\frac\pi8$ & 2.5 & 0.94 &  2.04 & 1.95 \\
 1.0 - 1.8 & 0-$\frac\pi8$ & 2.7 & 0.96 &  2.02 & 1.90
\\
\hline
\end{tabular}

\newpage

\begin{figure}[h]
\centerline{\epsfig{file=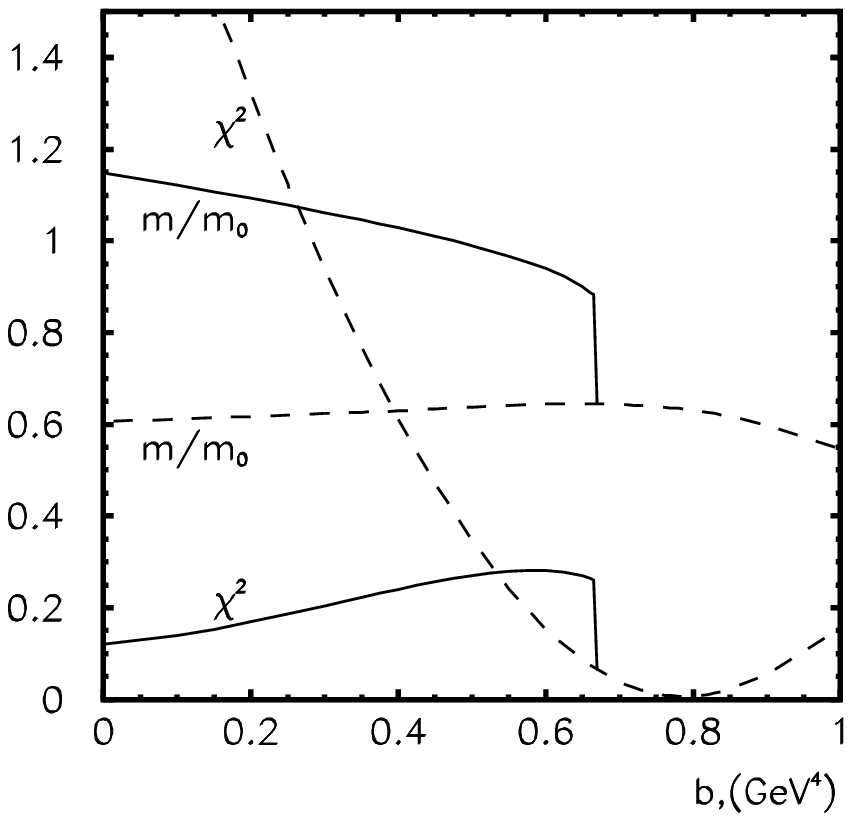,width=8cm}}
\caption{}
\label{fi2}
\end{figure}

\begin{figure}[h]
\centerline{\epsfig{file=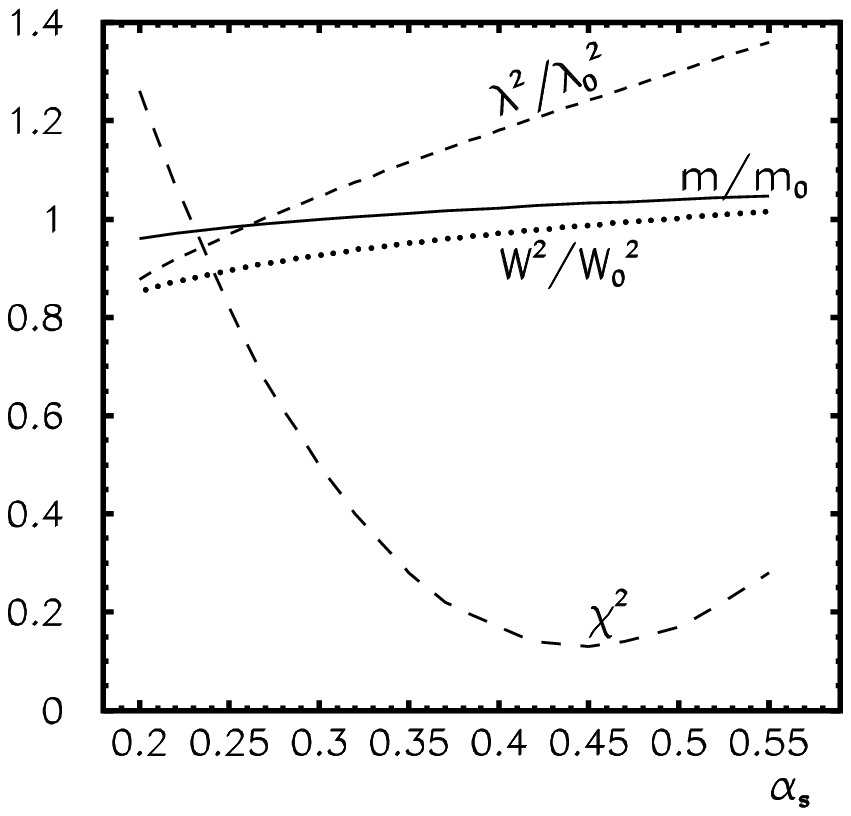,width=8cm}}
\caption{}
\label{fi1}
\end{figure}

\begin{figure}[h]
\centerline{\epsfig{file=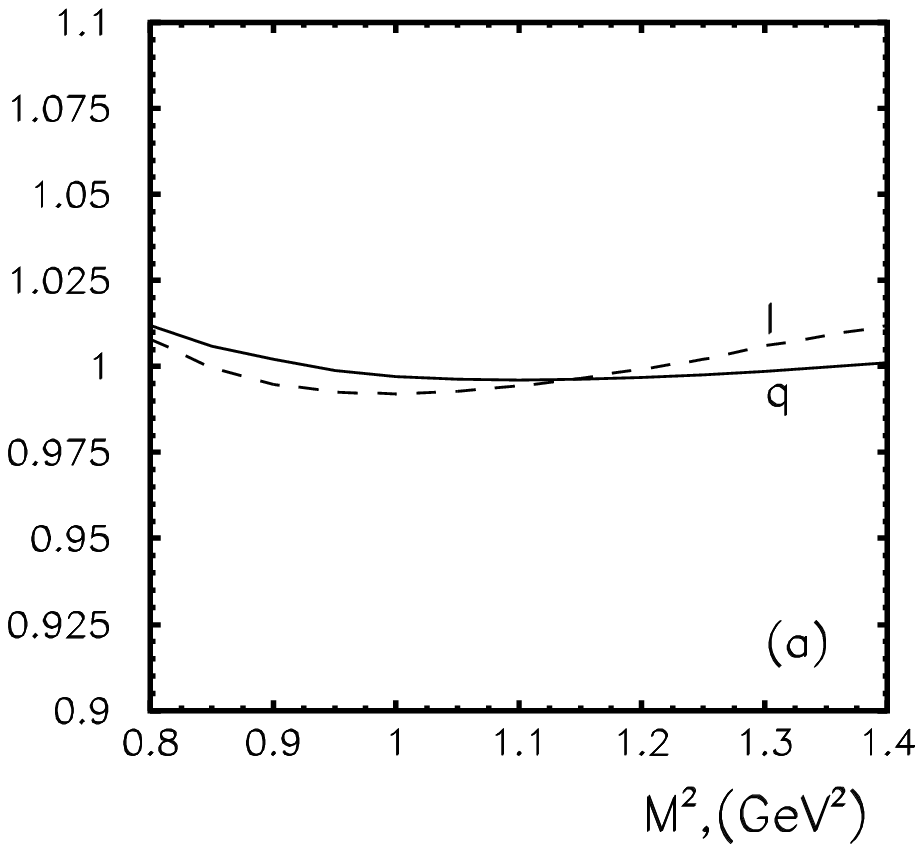,width=7cm}
\epsfig{file=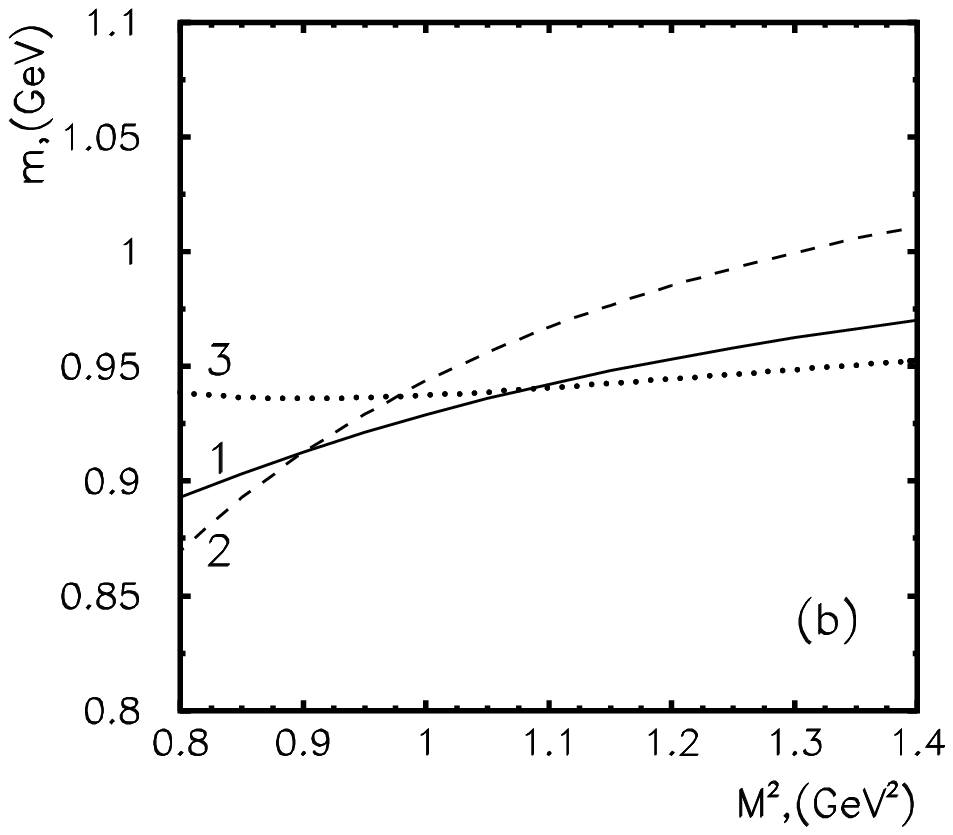,width=7cm}}
\caption{}
\label{fi3}
\end{figure}

\begin{figure}[h]
\centerline{\epsfig{file=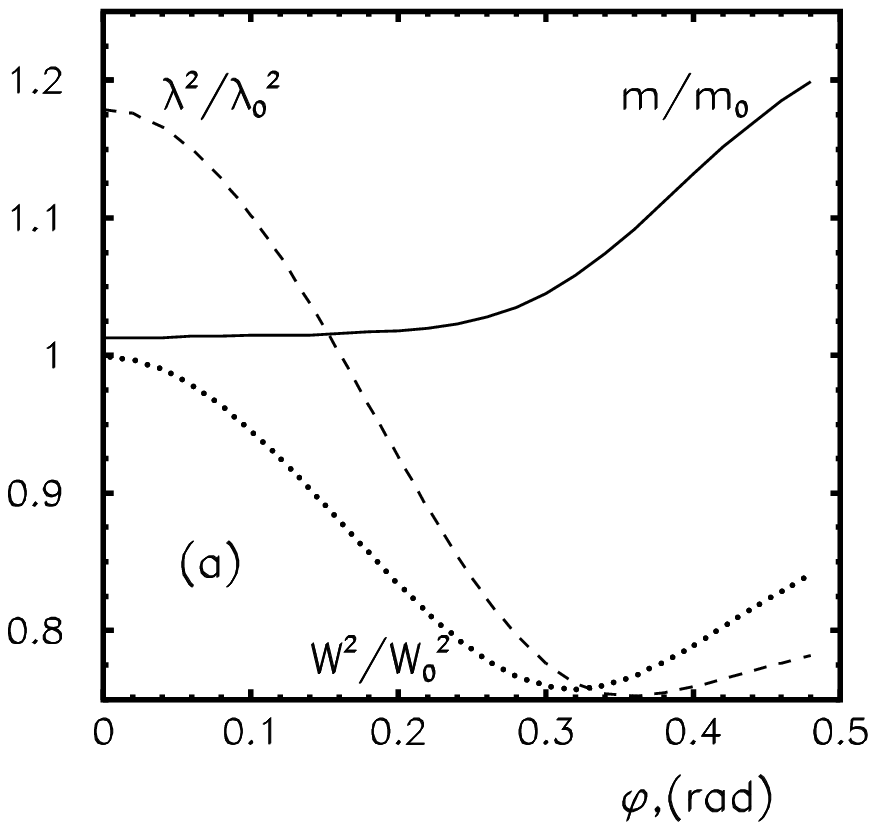,width=7cm}
\epsfig{file=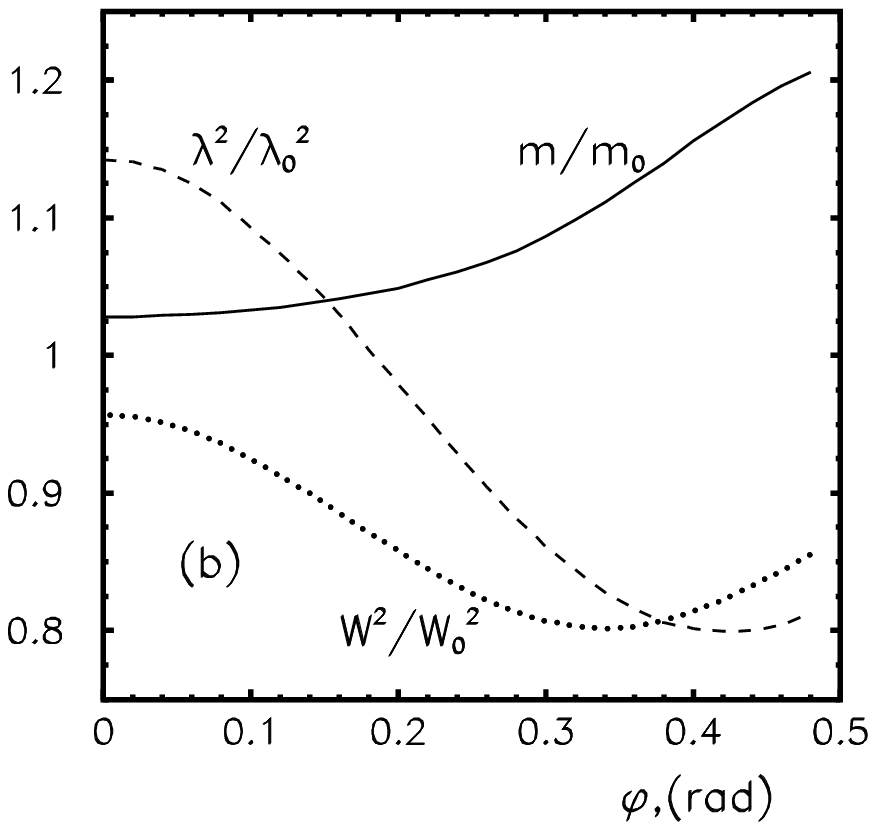,width=7cm}}
\caption{}
\label{fi4}
\end{figure}

\begin{figure}[h]
\centerline{\epsfig{file=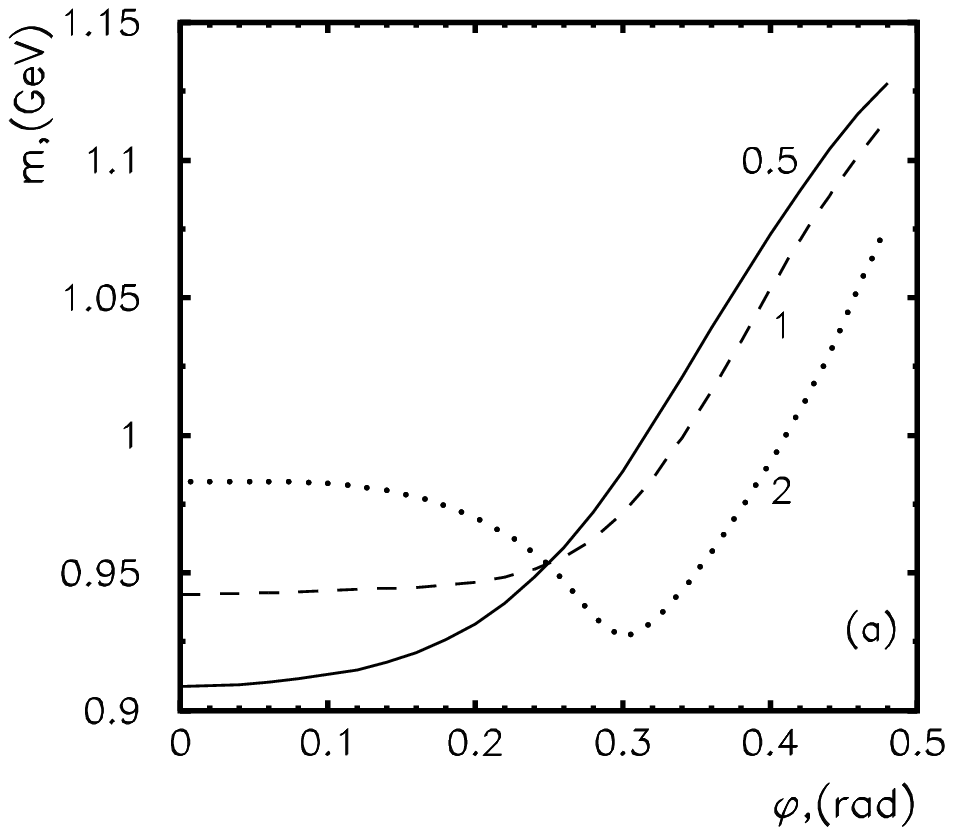,width=7cm}}
\centerline{\epsfig{file=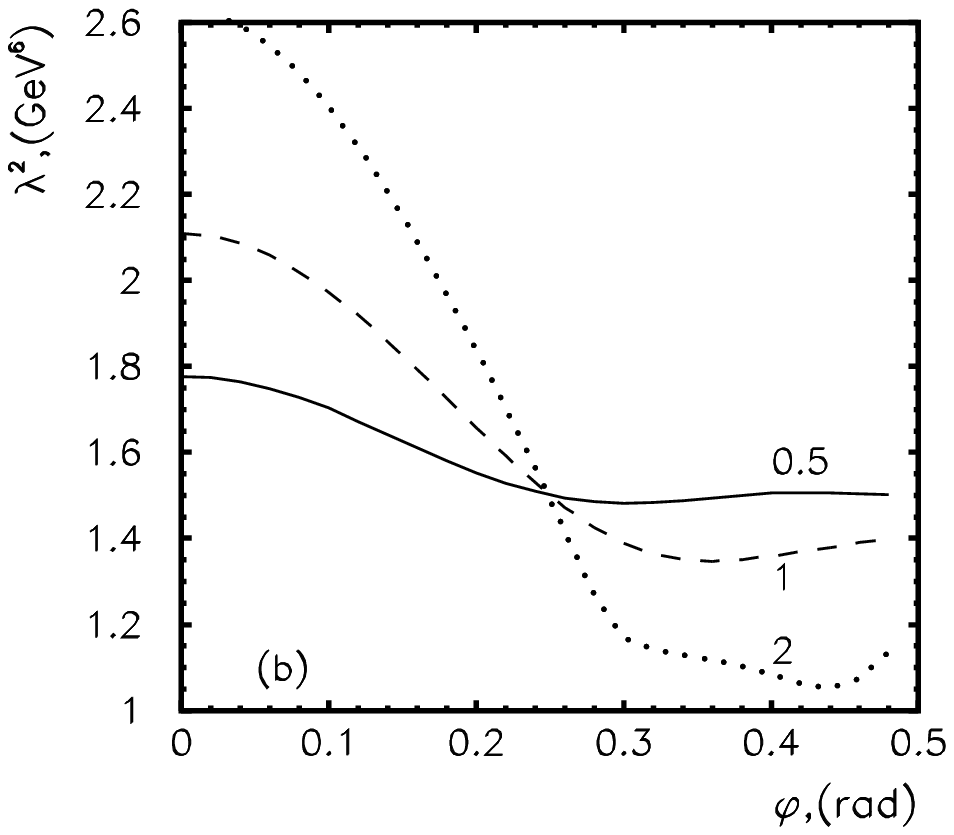,width=7cm}
\epsfig{file=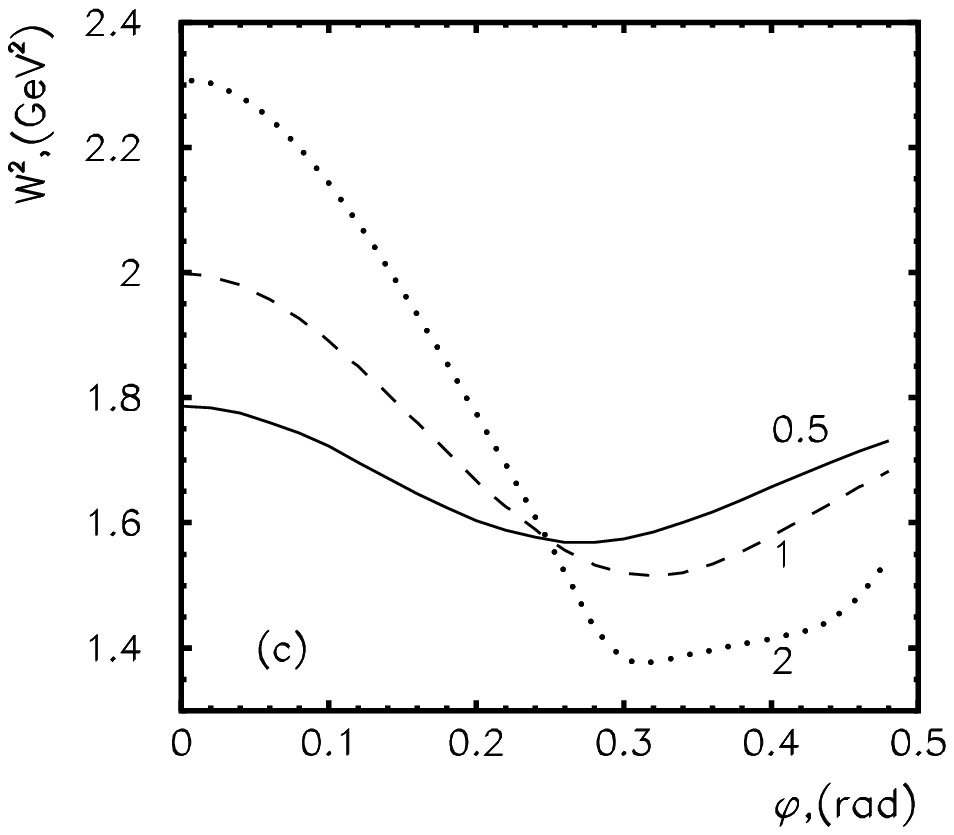,width=7cm}}
\caption{}
\label{fi6}
\end{figure}

\begin{figure}[h]
\centerline{\epsfig{file=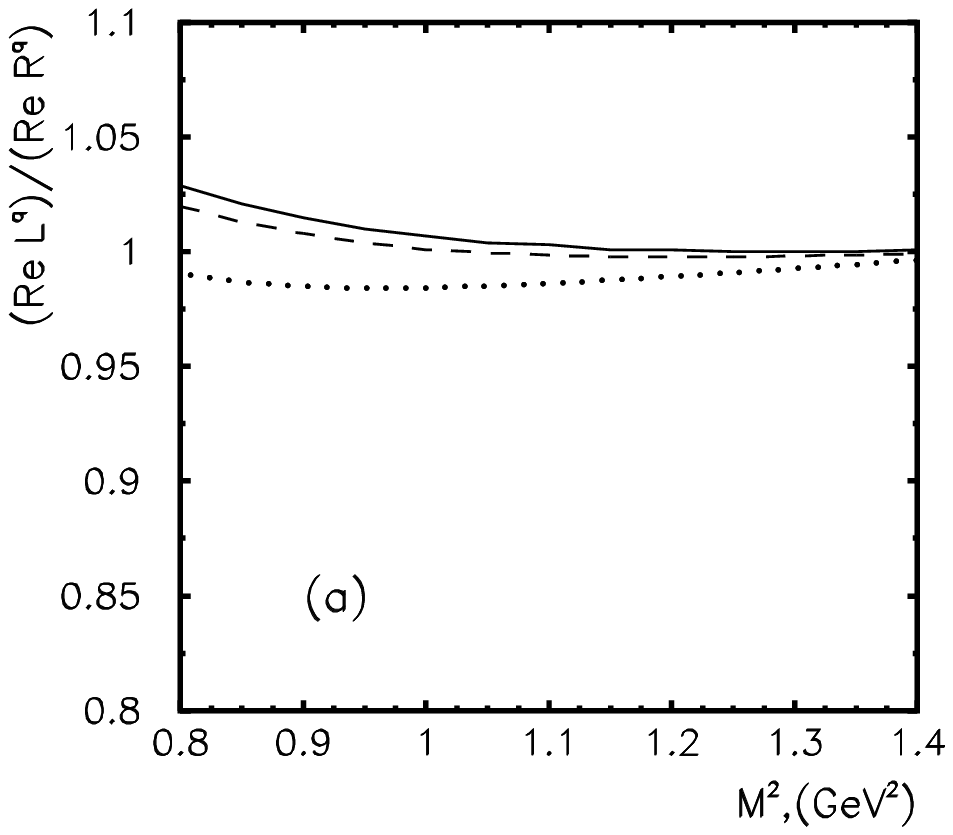,width=7cm}
\epsfig{file=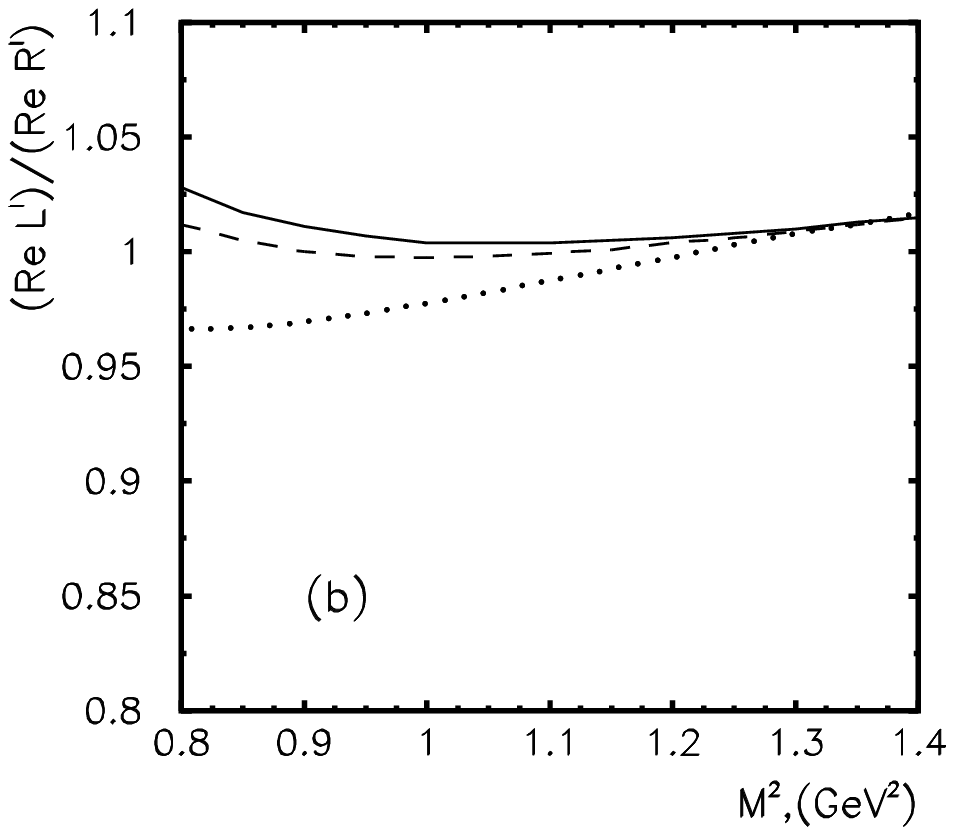,width=7cm}}
\caption{}\label{fi8}\end{figure}

\begin{figure}[h]
\centerline{\epsfig{file=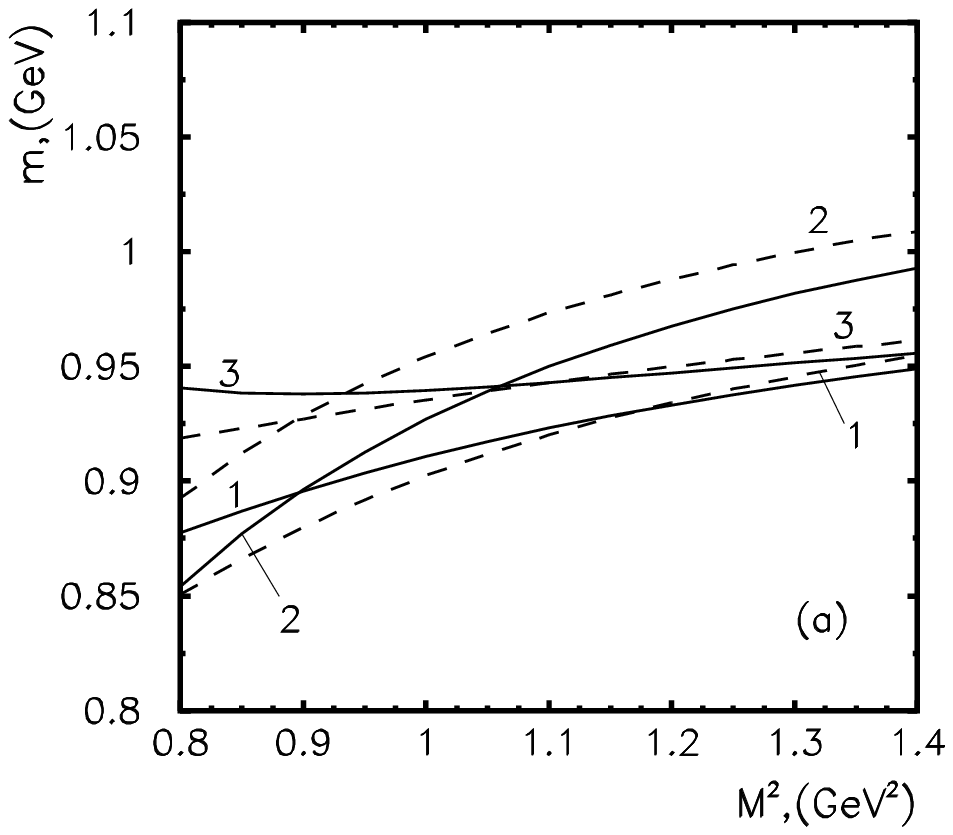,width=7cm}
\epsfig{file=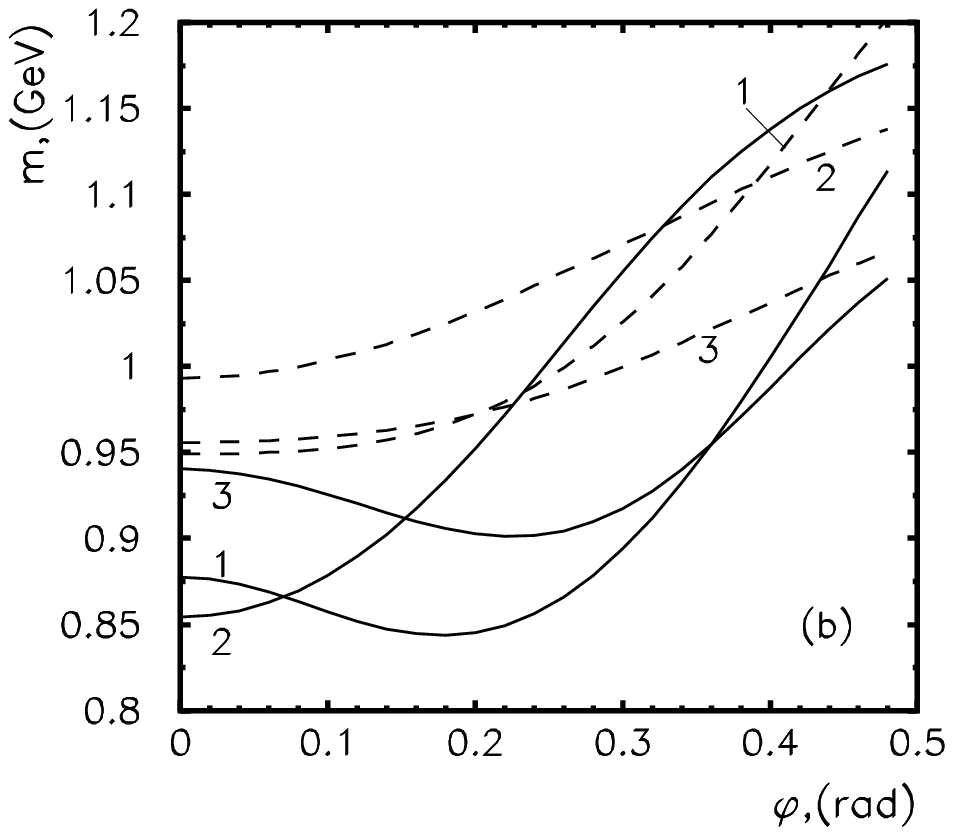,width=7cm}}
\caption{}
\label{fi8c}
\end{figure}

\begin{figure}[h]
\centerline{\epsfig{file=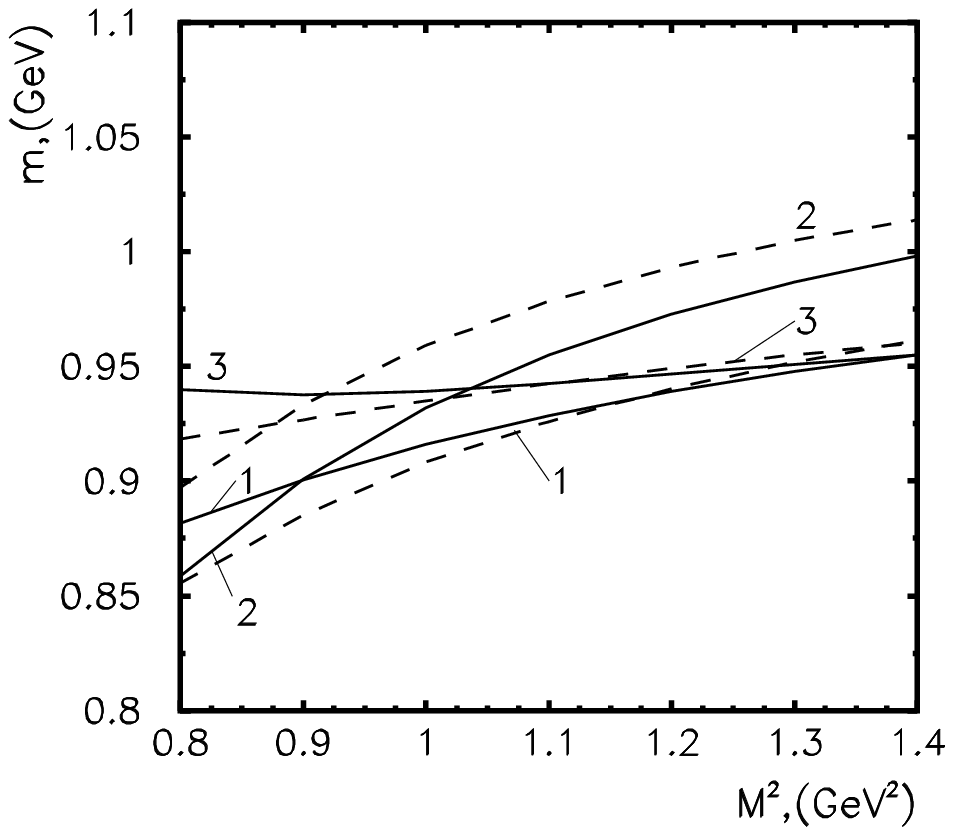,width=8cm}}
\caption{}
\label{fi9}
\end{figure}

\end{document}